\newcommand{\mb}[1]{\ifmmode#1\else\mbox{$#1$}\fi}

\newcommand\al{\mb{\alpha}}

\newcommand\ga{\mb{\gamma}}
\newcommand\de{\mb{\delta}}

\newcommand\De{\mb{\Delta}}

\newcommand{\beq}{\begin{equation}}
\newcommand{\eeq}{\end{equation}}
\newcommand{\nn}{\nonumber}
\newcommand{\bea}{\begin{eqnarray}}
\newcommand{\eea}{\end{eqnarray}}

\newcommand{\x}{\mb{\times}}

\newcommand{\Ad}{{\rm Ad}}
\newcommand{\gsim}
{\raise.3ex\hbox{$\;>$\kern-.75em\lower1ex\hbox{$\sim$}$\:$}}
\newcommand{\lsim}
{\raise.3ex\hbox{$\;<$\kern-.75em\lower1ex\hbox{$\sim$}$\:$}}

\newcommand{\ts}{\textstyle}
\newcommand{\half}{{\ts \frac{1}{2}}}
\newcommand{\third}{{\ts \frac{1}{3}}}
\newcommand{\twothird}{{\ts \frac{2}{3}}}

\documentstyle[prl,aps,twocolumn,floats]{revtex}

\begin{document}


\twocolumn[\hsize\textwidth\columnwidth\hsize\csname @twocolumnfalse\endcsname
\title{Gauge Unification in the Dual Standard Model}
\author{Nathan\  F.\  Lepora}
\address{Department of Applied Mathematics and Theoretical Physics,
Cambridge University, England}  
\date{October 27, 1999}
\maketitle

\begin{abstract}
We calculate the gauge couplings in the dual standard model. These
values are consistent with an associated GeV mass scale, and predict
the weak mixing angle to be ${\sin^2 \theta_w} (M_Z) \sim 0.22$.\\  

\ \\
\end{abstract}
]


The standard model fermions have an intricate representation structure
under the colour, weak isospin and hypercharge symmetry groups. The
five 
basic multiplets (replicated in three generations) divide into leptons
and quarks corresponding to trivial and fundamental representations of
the colour symmetry. These leptons and quarks subdivide further
corresponding to the trivial or fundamental representations of
weak isospin,
with this division coinciding with left and right parity eigenstates.

Currently, the {\em only} explanation for such a structure is the dual
standard model of Vachaspati~\cite{vacha}. Here the standard model
fermions are associated with monopoles originating from the symmetry
breaking of 
a unified $SU(5)$ gauge theory to the standard model gauge
symmetry. The representation structure, and hence interaction, of
these monopoles is in exact agreement with the spectrum of fermions in
the standard model. 

In addition, other properties such as the spin~\cite{vach1}, and the
number of 
generations can be consistently included within this
framework\cite{Liu:1997cc}. Its structure may also be related to
confinement within QCD~\cite{gold,vachb}.

In this letter we calculate the colour, weak isospin and hypercharge
gauge coupling constants of these $SU(5)$ monopoles. Essentially our
calculation compares the gauge transformation 
properties of the monopoles with the associated fermions. This
uniquely specifies the colour, weak isospin and hypercharge
gauge coupling constants in terms of the unified $SU(5)$ coupling. 
We find these gauge couplings to be consistent with the experimental
values.  

We begin by summarising some of the main features of the dual standard
model~\cite{vacha,vachb}. The model originates with a breaking
of $SU(5)$ gauge symmetry
\bea
\label{dualsb}
SU(5) \rightarrow S(U(3) &\x& U(2)) \nn \\
&=& [SU(3) \x SU(2) \x U(1)]/{\bf Z}_6 
\eea
and has a  monopole spectrum corresponding to the homotopy classes
\bea
\label{homotopy}
\pi_2 \left( \frac{SU(5)}{S(U(3) \x U(2))} \right) 
&\cong& \pi_1(S(U(3) \x U(2))) \nn \\
&=& {\bf Z}_6 \x {\bf Z}.
\eea
Here ${\bf Z}$ defines the degree of the homotopy class, whilst
\beq
{\bf Z}_6={\bf Z}_3 \x {\bf Z}_2=\{e^{2i\pi/3}, e^{-2i\pi/3}, 1\}
\x \{-1, 1\}
\eeq
represents second homotopy classes of same degree. 

The monopoles spectrum is built up from bound states of embedded
$SU(2) \rightarrow U(1)$ fundamental monopoles,
\bea
\label{embedding}
SU(5) &\rightarrow& S(U(3) \x U(2)) \nn \\
\cup  \hspace{1em} &\ &\  \cup \\
SU(2) &\rightarrow& U(1), \nn
\eea
and correspond to the $(e^{2i\pi/3}, -1)$ homotopy class of ${\bf
Z}_6$. Gardner and Harvey~\cite{gard84} show that these fundamental
monopoles combine to form stable bound states for a natural range of
model parameters. Labelling the bound states by their asymptotic
magnetic fields  
\beq
B^k \sim \frac{\hat{r}^k}{r^2}  {Q},
\eeq
defines an associated magnetic charge
\beq
\label{Q}
{Q}=\frac{1}{g_u} \left( q_{\rm C}T_{\rm C} + q_{\rm I}T_{\rm I} +
q_{\rm Y}T_{\rm Y} \right), 
\eeq
where $T_{\rm C}, T_{\rm I}$ and $T_{\rm Y}$ are suitably normalised
elements of the Lie algebras $su(3), su(2)$ and $u(1)$. The
coefficient $1/g_u$ relates to the unified $SU(5)$ gauge coupling
$g_u$. 

The magnetic charges are determined by associating them with the
corresponding homotopy classes in Eq.~(\ref{homotopy}). They define a
subgroup  
\beq
U(1)_{Q} = \exp({\bf R} Q) \subset S(U(3)\x U(2)),
\eeq
normalised by 
\beq
\exp(2\pi g_u  {Q})=1.
\eeq
This subgroup represents a typical element of the
associated ${\bf Z}_6$ homotopy class of the monopole. Using
generators 
\bea
\label{gena}
T_{\rm C}&=& i\ 
{\rm diag}(-\third, -\third, \twothird,0,0),\\ 
\label{genb}
T_{\rm I}&=& i\  {\rm diag}(0,0,0, 1, -1),\\
\label{genc}
T_{\rm Y}&=& i\  {\rm diag}(1, 1, 1, -{\ts \frac{3}{2}}, -{\ts
\frac{3}{2}})  
\eea
leads to the following pattern for the monopole spectrum:
\begin{center}
\begin{tabular}{|c|ccc|ccc|}
\hline
\  & $q_{\rm C}$ & $q_{\rm I}$ & $q_{\rm Y}$ & $d_{\rm C}$ &
$d_{\rm I}$ & $d_{\rm Y}$ \\ 
\hline
$(e^{2i\pi/3}, -1)$ & 1 & 1/2 & 1/3 & 3 & 2 & 1\\ 
$(e^{-2i\pi/3}, 1)$ & -1 & 0 & 2/3 & 3 & 0 & 1 \\
$(1, -1)$ & 0 & 1/2 & 1 & 0 & 2 & 1 \\
$(e^{2i\pi/3}, 1)$ & 1 & 0 & 4/3 & 3 & 0 & 1\\
$(e^{-2i\pi/3}, -1)$ & - & - & - & - & - & - \\
$(1, 1)$ & 0 & 0 & 2 & 0 & 0 & 1 \\
\hline
\end{tabular}
\end{center}
It should be noted that we have chosen a slightly different
normalisation from~\cite{vacha,vachb}. This is to agree with the
standard particle physics charge normalisations.

Degeneracies $d_{\rm C}$, $d_{\rm I}$ and $d_{\rm Y}$ of the monopole 
embeddings corresponding to the same homotopy class have also been
included. These arise from the degeneracy of suitable generators
\bea
T^r_{\rm C}&=& i\ 
{\rm diag}(+\twothird, -\third, -\third,0,0),\\ 
T^g_{\rm C}&=& i\ 
{\rm diag}(-\third, +\twothird, -\third,0,0),\\ 
T^b_{\rm C}&=& i\ 
{\rm diag}(-\third, -\third, +\twothird,0,0)
\eea
for $X_{\rm C}$ and 
\beq
T_{\rm I}^\pm = \pm T_{\rm I}
\eeq
for $T_{\rm I}$. This indicates the monopoles form
representations of $SU(3)_{\rm C}$, $SU(2)_{\rm I}$ and $U(1)_{\rm Y}$
with the corresponding dimension. Namely the fundamental
representations. 

The above arguments strongly imply that the long range interactions of
these monopoles is associated with that of a particle with gauge
interactions specified by the fundamental representations of the
colour, weak isospin and hypercharge symmetry groups. This particle has
the corresponding charges $q_{\rm C}$, $q_{\rm I}$ and $q_{\rm Y}$, and
its current $J^\mu_{\rm mon}$ couples to the gauge fields as
\beq
\label{coupling}
[g_{\rm C} q_{\rm C} A^\mu_{\rm C} + g_{\rm I} q_{\rm I} A^\mu_{\rm I}
+ g_{\rm Y} q_{\rm Y} A^\mu_{\rm Y}]\, J^\mu_{\rm mon},
\eeq
with $g_{\rm C}$, $g_{\rm I}$ and $g_{\rm Y}$ representing the
respective gauge couplings. Such a spectrum of charges and
interactions is completely in accord 
with the spectrum of fermions in the standard model, with the
identification:
\bea
(e^{2i\pi/3}, -1) &\leftrightarrow& (u,d)_L \nn \\
(e^{-2i\pi/3}, 1) &\leftrightarrow&  \bar{d}_L \nn \\
(1, -1)\ \ \  &\leftrightarrow& (\bar\nu,\bar e)_R  \\
(e^{2i\pi/3}, 1)\  &\leftrightarrow& u_R \nn \\
(1, 1)\ \ \ \  &\leftrightarrow& \bar{e}_L \nn
\eea
The corresponding fermionic anti-particles are associated with the
anti-monopoles. 

On the question of duality, both the residual symmetry group $S(U(3)\x U(2))$
and its dual $S(U(3)\x U(2))^v=SU(3)\x SU(2) \x U(1)$ have the same derived
representation, because their local structure is equivalent. Thus in both
cases the action of their associated gauge fields on particle
representations are the same.

The main point of this work is to show that as well as predicting the
spectrum and properties of fermions in the standard model, the dual
standard model also predicts the corresponding colour, weak isospin
and hypercharge gauge couplings. We shall determine these from comparing the  
gauge couplings of the monopole currents to the corresponding expressions for
their associated fermions. We shall also give an alternative, but equivalent,
argument from the gauge transformation properties of the monopoles. 

One may see simply that three different gauge couplings 
arise in Eq.~(\ref{coupling}) by considering the normalisation of the 
monopole charge generators in 
Eqs.~(\ref{gena}, \ref{genb}, \ref{genc}). These generators $T_{\rm
C}$, $T_{\rm I}$ and $T_{\rm Y}$ are normalised to the topology of
$S(U(3) \x U(2))$. However the gauge fields of $SU(5)$ theory are
normalised differently. In the minimal coupling the components of the
gauge fields are written
\beq
\label{su5a}
D^\mu = \partial^\mu + g_u A^\mu_a \hat{T}_a,
\eeq
with the $SU(5)$-basis $\{\hat{T}_a\}$ orthonormal with respect to the
inner product
\beq
\label{su5b}
{\rm tr}(\hat{T}_a \hat{T}_b) = \half \de_{ab}.
\eeq
The difference between these normalisations will produce overall
scales associated with the gauge couplings.

We shall illustrate the importance of normalisation with the coupling
of standard model fermions to their gauge fields. A fermion $f$ couples to
gauge fields through its current $j^\mu$. In particular we shall consider the
neutral current components
\beq
j^\mu_{\rm C} = \bar{f} \ga^\mu {X}_{\rm C} f,\ \ 
j^\mu_{\rm I} = \bar{f} \ga^\mu {X}_{\rm I} f,\ \ 
j^\mu_{\rm Y} = \bar{f} \ga^\mu {X}_{\rm Y} f,
\eeq
where the standard generators are
\bea
 X_{\rm C} &=& i\:{\rm diag}(1,1,-2),\\
 X_{\rm I} &=& i\:{\rm diag}(1,-1),\\
 X_{\rm Y} &=& i\:{\rm diag}(1,1).
\eea
It is important to take a standard $su(3)$, $su(2)$ and $u(1)$
normalisation 
\beq
\label{fermnorm}
{\rm tr}(\hat X_{\rm C}^2)=
{\rm tr}(\hat X_{\rm I}^2)=
{\rm tr}(\hat X_{\rm Y}^2)=\half,
\eeq
which we will explicitly include by considering
\beq
\hat X_{\rm C}= {\ts {\frac{1}{\sqrt{12}}}} X_{\rm C},\ \ 
\hat X_{\rm I}= \half X_{\rm I},\ \ 
\hat X_{\rm Y}= \half X_{\rm Y}.
\eeq
Then a fermion with colour charge $q_{\rm C}$, weak isospin $q_{\rm I}$,
and weak hypercharge $q_{\rm Y}$ has a gauge-current coupling of the form
\beq
\label{ferm}
{\ts {\frac{1}{\sqrt{12}}}} 
g_{\rm C} q_{\rm C} A_{\rm C}^{\mu} j^\mu_{\rm C} +
\half g_{\rm I} q_{\rm I} A_{\rm I}^{\mu} j^\mu_{\rm I} +
\half g_{\rm Y} q_{\rm Y} A_{\rm Y}^{\mu} j^\mu_{\rm Y}.
\eeq

Now we shall consider the coupling of the corresponding monopoles to their
gauge fields. From the above arguments leading to Eq.~(\ref{coupling}) we may
take the monopoles as coupling to $S(U(3)\x U(2))$ gauge fields through their
associated currents. In particular we shall consider three neutral components
of the monopole current, 
$J^\mu_{\rm C}$, $J^\mu_{\rm I}$, and $J^\mu_{\rm Y}$. These are associated
with generators
\bea
T_{\rm C}&=& i\ 
{\rm diag}(-\third, -\third, \twothird,0,0),\\ 
T_{\rm I}&=& i\  {\rm diag}(0,0,0, 1, -1),\\
T_{\rm Y}&=& i\  {\rm diag}(1, 1, 1, -{\ts \frac{3}{2}}, -{\ts
\frac{3}{2}}). 
\eea
We shall take a standard $su(5)$ normalisation
\beq
\label{monnorm}
{\rm tr}(\hat T_{\rm C}^2)=
{\rm tr}(\hat T_{\rm I}^2)=
{\rm tr}(\hat T_{\rm Y}^2)=\half,
\eeq
which will be explicitly included by considering
\beq
\hat T_{\rm C}= {\ts {\frac{\sqrt{3}}{2}}} T_{\rm C},\ \ 
\hat T_{\rm I}= \half T_{\rm I},\ \ 
\hat T_{\rm Y}= {\ts {\frac{1}{\sqrt{15}}}} T_{\rm Y}.
\eeq
Then a monopole with colour charge $q_{\rm C}$, weak isospin $q_{\rm I}$,
and weak hypercharge $q_{\rm Y}$ has a gauge-current coupling of the form
\beq
\label{mon}
g_{\rm u}[ {\ts {\frac{\sqrt{3}}{2}}}
q_{\rm C} A_{\rm C}^{\mu} J^\mu_{\rm C} +
\half q_{\rm I} A_{\rm I}^{\mu} J^\mu_{\rm I} +
{\ts {\frac{1}{\sqrt{15}}}} q_{\rm Y} A_{\rm Y}^{\mu} J^\mu_{\rm Y}],
\eeq
where the gauge fields are considered as components of the $SU(5)$
gauge field, with a unified coupling $g_u$.

By associating these monopoles with standard model fermions we associate each
monopole current $J^\mu$ with a corresponding fermion current $j^\mu$. We also
associate the corresponding gauge fields. Thus the monopole-gauge coupling of
Eq.~(\ref{mon}) and fermion-gauge couplings of Eq.~(\ref{ferm}) are
identified. Comparison of the respective coefficients then gives
\beq
\label{res}
g_{\rm C} = 3 g_u,\ \ \ 
g_{\rm I} =  g_u,\ \ \ 
g_{\rm Y} = \frac{2}{\sqrt{15}} g_u,
\eeq
which predicts the following ratios:
\beq
\label{pred}
\frac{g_{\rm C}}{g_{\rm I}} = 3,\ \ \frac{g_{\rm Y}}{g_{\rm I}} =
\frac{2}{\sqrt{15}}. 
\eeq
Such values represent a specific prediction of the dual standard model
and are completely characteristic of it. 

The above relation may also be seen from the explicit transformation
properties of the monopoles. Recall that the fundamental monopoles are
embedded $SU(2) \rightarrow U(1)$ monopoles, described by
Eq.~(\ref{embedding}), with  magnetic fields $B^k$ corresponding to the
embedding. Rigid (or global) monopole gauge transformations that respect $B^k
\in su(3)_{\rm C} \oplus su(2)_I \oplus u(1)_{\rm Y}$ transform  
\beq
\label{mon-gauge}
B^k \mapsto \Ad(h) B^k
\eeq
under the adjoint action of $h \in S(U(3) \x U(2))$. Correspondingly
the $su(2)$ embedding transforms under
\beq
su(2) \mapsto \Ad(h) su(2),
\eeq
so that $Q$ transforms appropriately. 

Consider a rigid gauge transformation of the embedded monopole, with the
generators normalised as in Eq.~(\ref{monnorm})
\beq
B^k \mapsto \Ad[\exp(g_u(\hat{T}_{\rm C} \theta_{\rm C} +
\hat{T}_{\rm I} \theta_{\rm I} + \hat{T}_{\rm Y} \theta_{\rm Y}))]
B^k. 
\eeq
Those taking $B^k \mapsto B^k$ are thus 
\beq
\theta_{\rm C} = \frac{2}{\sqrt 3} \frac{2\pi}{g_u} n_{\rm C},\ \ 
\theta_{\rm I} = 2 \frac{2\pi}{g_u} n_{\rm I},\ \ 
\theta_{\rm Y} = \sqrt{15} \frac{2\pi}{g_u} n_{\rm Y},
\label{theta}
\eeq
with each $n \in {\bf N}$.

Now we associate the above transformation with an analagous rigid gauge
transformations on a fermion $f$ in the fundamental representation of
$SU(3)_{\rm C}\x SU(2)_{\rm I}\x U(1)_{\rm Y}/{\bf Z}_6$  
\beq
f \mapsto e^{g_{\rm Y}\bar\theta_{\rm Y}\hat{X}_{\rm Y}}
 \exp(g_{\rm C} \bar\theta_{\rm C} \hat X_{\rm C}) \ f\ 
\exp(g_{\rm I} \bar\theta_{\rm I} \hat X_{\rm I}), 
\eeq
with $g_{\rm C}$, $g_{\rm I}$ and $g_{\rm Y}$ the colour, weak isospin and
hypercharge gauge 
couplings. Those rigid gauge transformation that take $f \mapsto f$ are thus
\beq
\bar\theta_{\rm C} = \sqrt{12}\frac{2\pi}{g_{\rm C}} n_{\rm C},\ \ 
\bar\theta_{\rm I} = 2\frac{2\pi}{g_{\rm I}} n_{\rm I},\ \ 
\bar\theta_{\rm Y} = 2\frac{2\pi}{g_{\rm Y}} n_{\rm Y},
\label{thetabar}
\eeq
with each $n \in {\bf N}$.

Equating monopole and fermion gauge transformation identifies each
$\theta$ and $\bar\theta$ in Eqs.~(\ref{theta}) and (\ref{thetabar}).
This again gives the ratios found in Eq.~(\ref{res}).

These predictions are compared to the running gauge couplings through
the following plot. The strong coupling is taken from a three loop
calculation normalised to $g_{\rm C}(M_Z)=1.213$. The hypercharge and weak
isospin are taken from one loop expressions normalised to
$g_{\rm I}(M_Z)=0.661$ and $g_{\rm Y}(M_Z)=0.354$.
\begin{figure}[h]
\vspace{2.7in}
\includegraphics{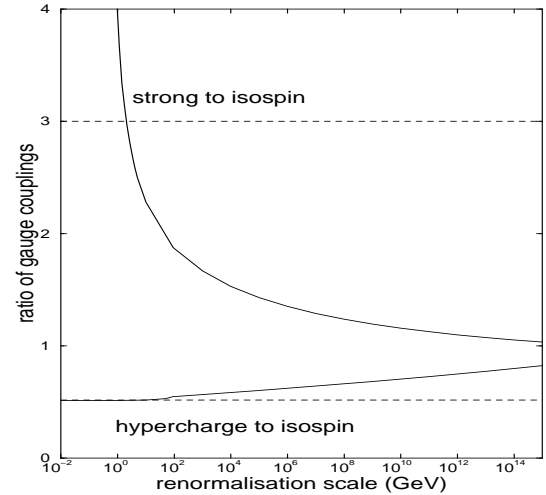}
\caption{$g_{\rm C}/g_{\rm I}$ and $g_{\rm Y}/g_{\rm I}$ plotted against
renormalisation scale 
$\mu$. Dual standard model predicted values are also included.} 
\label{fig}
\end{figure}

We shall make a couple of comments about the running of the gauge
couplings in the standard model. Firstly, 
around $10^{15}-10^{18}$GeV, when $g_{\rm C}/g_{\rm I} \sim 1$ then also
$g_{\rm Y}/g_{\rm I} \sim \sqrt{3/5}$, as required for grand unification.
Secondly, $g_{\rm Y}/g_{\rm I}$ runs below the Z-mass from the
running of the fine structure constant $\al$. Its form may be
estimated by the following relation~\cite{Erler:1998ig}  
\beq
\label{running}
M_W = \frac{A_0}{\sin \theta_w(1-\De r)^{1/2}},\ \ \ 
A_0 = (\pi\al /\sqrt 2 G_F)^{1/2},
\eeq
with $\De r$ representing the radiative corrections. Its component
from the running of $\al$ is $\De r_0(\mu) = (1-\al/\al(\mu))$.

The conclusion from fig.~(\ref{fig}) is that the dual standard model
is associated with a mass scale of around a few GeV. At that scale
the running couplings take the values of {\em both} of our theoretical
predictions in Eq.~(\ref{pred}). 

To illustrate the accuracy of the fit in fig.~(\ref{fig}) we shall
calculate a prediction for $\sin^2 \theta_w
(M_Z)$ using only the running of the strong coupling and
Eq.~(\ref{pred}). Firstly observe that $g_{\rm C}/g_{\rm I}=3$ is satisfied at
around a few GeV. Then Eq.~(\ref{pred}) implies that $\sin^2 \theta_w= 4/19$
at the same scale. Using Eq.~(\ref{running}), we predict
\beq
\label{theta_w}
\sin^2 \theta_w (M_Z) \sim \frac{\al(M_Z)}{\al}\sin^2 \theta_w (0)
\sim 0.22.
\eeq
The experimental value is $\sin^2 \theta_w (M_Z) = 0.2230\pm 0.0004$.


The above is the conclusion of this work. We considered the long range
interactions of monopoles in the dual standard model to be given via the
colour, isospin and hypercharge gauge fields. Then, by an appropriate
appreciation of the associated normalisations of the gauge fields, we derived
relations between the colour, isospin and hypercharge gauge fields at the
scale of monopole unification. These values were found to be consistent with
standard model gauge couplings, and the degree of their consistency may be
appreciated through the prediction of $\sin^2 \theta_w (M_Z)$ in
Eq.~(\ref{theta_w}). 

We think that the above results should be appreciated independently of any
interpretation placed on them. It is our aim to present the above mathematics
as self consistent, and arising through the geometric structure of the
monopoles occuring in Georgi-Glashow $SU(5)$ theory. However, since the
agreement is so precise one must speculate somewhat on the fundamental
structure that gives rise to this agreement. This is the subject of the rest
of this letter, although it should be appreciated that the results we have
presented thus far should be considered independently of the following
discussion. Indeed, all of the following interpretations may be incorrect. 

A first interpretation is coincidence. One may achieve no further implication
from such an interpretation.

A second, conventional, interpretation is the relations are arising from
some duality between the standard model fermions and the non-perturbative
features of Georgi-Glashow $SU(5)$ symmetry breaking. Presumably such duality
gives rise to consistency relation in the gauge coupling constants of the
standard model. The existence of such a duality is most likely to arise within
a string, or some membrane theory, where examples of analogous dualities are
known. In this context the relations between the standard model gauge
couplings constants that we have derived could be interpreted as a direct
low energy implication of the fundamental string or membrane unification
picture.  

We believe to give weight to such a proposal a specific duality of the
fundamental unification would need to be found. Such a question is an
interesting proposal, and in our opinion should be investigated further.
However, it is beyond the scope of the present paper to discuss this further.

A third, and more unconventional, interpretation is the observed
fermions of the standard model really are monopoles. They are formed at
monopole unification, where the fundamental gauge symmetries
unify. Above this scale there are no fermions, and matter exists
solely in the form of fundamental fields that make up the unified
gauge theory. 

The necessary, and dramatic feature of such an interpretation is that
gauge unification occurs at a plasma temperature of around a few GeV. Clearly
this feature seems problematic, indeed is completely contradictory to the
present viewpoint on unification. However, such plasma temperatures have not
yet been reached and, in our opinion, until they have the consequences of such
a suggestion should be explored.

It should be noted that in one context unification at a few GeV is
desirable. Typical masses of the monopoles are of the unification scale, which
is a fairly typical mass scale of the standard model fermions, somewhere
between the charmed and bottom quark masses. Thus at least this mass scale is
consistent with the fermion masses. In fact, in this context, if monopole
unification scale were much higher it would be difficult to reconcile with the
observed fermion masses. 

In conclusion we have examined the gauge couplings in the dual
standard model. This model represents a theoretically well motivated
explanation of the spectrum and interaction of the standard model
fermions. We have shown that it predicts
$g_{\rm C}/g_{\rm I}=3$ and $g_{\rm Y}/g_{\rm I}=2/\sqrt{15}$, values that are
consistent with the standard model gauge couplings at a renormalisation scale
around a few GeV.

{\it
I acknowledge King's College, Cambridge for a junior research
fellowship and thank B.Allanach, P.Saffin and T.Vachaspati for their
valuable comments and advice. I also thank M.Ortiz for
discussions when this work was in an early state of development. 
}


\end{document}